\def\be{\begin{equation}}
\def\ee{\end{equation}}
\def\ba{\begin{array}}
\def\ea{\end{array}}

\def\Rb{{I\!\! R}}

\def\Cb{{\Bbb C}}

\documentstyle[12pt]{article}
\begin{document}
\input amssym.def
\setcounter{page}{1} \centerline{\Large\bf Equivalence
of Quantum States}
\vspace{2ex} \centerline{\Large\bf under Local
Unitary Transformations}
\vspace{4ex}
\begin{center}

Shao-Ming Fei$^{a,b}$
\footnote{e-mail: fei@uni-bonn.de}, ~Naihuan Jing$^{c, d}$
\footnote{e-mail: jing@math.ncsu.edu}

\vspace{2ex}
\begin{minipage}{5in}

{\small $~^{a}$ Department of Mathematics, Capital Normal
University, Beijing 100037}

{\small $~^{b}$ Institut f\"ur Angewandte Mathematik,
Universit\"at Bonn, D-53115}

{\small $~^{c}$ Department of Mathematics,
North Carolina State University,

~~~Raleigh, NC 27695}

{\small $~^{d}$ Department of Mathematics,
Hubei University, Wuhan, Hubei 430062}

\end{minipage}
\end{center}

\vskip 1 true cm
\parindent=18pt
\parskip=6pt
\begin{center}
\begin{minipage}{4.5in}
\vspace{3ex} \centerline{\large Abstract} \vspace{4ex}

In terms of the analysis of fixed point subgroup and tensor
decomposability of certain matrices, we study the equivalence of
of quantum bipartite mixed states under local unitary transformations.
For non-degenerate case an operational criterion for the
equivalence of two such mixed bipartite states under
local unitary transformations is presented.

\bigskip
\bigskip

PACS numbers: 03.67.-a, 02.20.Hj, 03.65.-w\vfill
\smallskip
MSC numbers: 94A15, 62B10\vfill
\smallskip

Key words: Local unitary transformation;
Quantum states; Tensor decomposable\vfill

\end{minipage}
\end{center}

\bigskip
\medskip
\bigskip

As one of the most striking features of
quantum phenomena \cite{1}, quantum entanglement
has been identified as a non-local resource for quantum
information processing such as quantum computation
\cite{DiVincenzo,book}, quantum teleportation
\cite{teleport}, dense coding
\cite{dense}, quantum cryptographic schemes
\cite{crypto1}£¬entanglement swapping \cite{4}, and remote
state preparation (RSP) \cite{5,6,7,8} etc.. As a matter
of fact, the degree of entanglement of two parts of a quantum system
remains invariant under local unitary transformations of these
parts. And two quantum states with the same
degree of entanglement (say, entanglement of formation
\cite{Bennett96a}) may be not equivalent under local unitary
transformations \cite{book}. It is important to classify
and characterize the quantum states in terms of local unitary
transformations. One way to deal with the problem is to find the
complete set of invariants of local unitary transformations. Two
states are equivalent under local unitary transformations if and
only if they have the same values of all these invariants. The
method developed in \cite{Rains,Grassl}, in principle, allows one
to compute all the invariants of local unitary transformations,
though in general it is not operational. In \cite{makhlin} the
invariants for general two-qubit systems are studied and a
complete set of 18 polynomial invariants is presented. In
\cite{linden,sud} three qubits states are also discussed. In
\cite{generic} a complete set of invariants is presented for
bipartite generic mixed states. In \cite{goswami} a complete set
of invariants under local unitary transformations is presented for
rank-2 and multiplicity free mixed states. In \cite{tri} the case
of tripartite is studied in detail and a complete set of
invariants is presented for a class of pure states.

Different approaches have their own advantages in dealing
with certain classes of quantum states.
In this letter, we investigate the equivalence problem in terms of
fixed point subgroup
and tensor decomposability of certain matrices. The problem is reduced
to verify whether a certain matrix is rank one or not. An operational
criterion is presented for the
equivalence of two non-degenerate mixed bipartite quantum states.

Let $H_1$ (resp. $H_2$) be an
$M$ (resp. $N$)-dimensional complex Hilbert space, with $\vert e_i\rangle$, $i=1,...,M$
(resp. $\vert f_j\rangle$, $j=1,...,N$),
as an orthonormal basis. A general pure state on $H_1\otimes H_2$ is of
the form
\begin{equation}\label{mmm}
\vert\Psi\rangle=\sum_{i=1}^M\sum_{j=1}^N a_{ij}\vert e_i\rangle \otimes
\vert f_j\rangle,~~~~~~a_{ij}\in\Cb
\end{equation}
with the normalization $\displaystyle\sum_{i=1}^M\sum_{j=1}^N
a_{ij}a_{ij}^\ast=1$ ($\ast$ denoting complex conjugation).

A bipartite quantum mixed state on $H_1\otimes H_2$
is described by a density matrix $\rho$
which can be decomposed according to its eigenvalues and eigenvectors:
$$
\rho=\sum_{i=1}^{MN}\lambda_i\vert\nu_i\rangle\langle\nu_i\vert,
$$
where $\lambda_i$ are the
eigenvalues and $\vert\nu_i\rangle$, $i=1,...,MN$, the
corresponding eigenvectors of the form (\ref{mmm}).

Two density matrices $\rho$ and $\rho^\prime$ are said to be equivalent
under local unitary transformations if there exist unitary operators
$U_1$ on $H_1$ and $U_2$ on $H_2$ such that
\be\label{eq}
\rho^\prime=(U_1\otimes U_2)\rho(U_1\otimes U_2)^\dag.
\ee

For a Hermitian matrix $A$ on $H_1\otimes H_2$, the set of commuting matrices $B$ such
that $AB=BA$ is called the \underline{commutant} of $A$, denoted
as $C(A)$. Obviously $C(A)$ is a subalgebra. In the following we
call the set of unitary matrices $U$ such that $UA=AU$ the
\underline{fixed point subgroup} of $A$, denoted as $C_U(A)$, which
is a subgroup of the unitary group of all unitary matrices.

{\sf [Definition].} If a matrix $V$ on $H_1\otimes H_2$ can be
written as $V=V_1\otimes V_2$ for
$V_1\in End(H_1)$, $V_2\in End(H_2)$, we say that $V$ is tensor
decomposable.

If two density matrices $\rho$ and $\rho^\prime$ are equivalent
under local unitary transformations, they must have the same set of
eigenvalues $\lambda_i$, $i=1,...,NM$. Let $X$ and $Y$ be the unitary
matrices that diagonalize $\rho$ and $\rho^\prime$ respectively,
\be\label{xy}
\rho=X\Lambda X^\dag,~~~~~\rho^\prime=Y\Lambda Y^\dag,
\ee
where $\Lambda=diag(\lambda_1,\lambda_2,...,\lambda_{MN})$.

{\sf [Lemma].} Let $G$ be the fixed point unitary subgroup
associated with $\rho$. Then $\rho^\prime$ is equivalent to $\rho$
under local unitary transformations if and only if the coset
$GXY^\dag$ contains a unitary tensor decomposable matrix.

{\sf[Proof].} Suppose $\rho^\prime=(U_1\otimes U_2)\rho(U_1\otimes U_2)^\dag$, then
$\rho^\prime=Y\Lambda Y^\dag=Y X^\dag \rho X Y^\dag$. Hence
$Y X^\dag \rho X Y^\dag=(U_1\otimes U_2)\rho(U_1\otimes U_2)^\dag$, or
$$
(U_1\otimes U_2)^\dag Y X^\dag \rho= \rho(U_1\otimes U_2)^\dag Y X^\dag.
$$
That is, $(U_1\otimes U_2)^\dag Y X^\dag\in C_U(\rho)$, and
$C_U(\rho)X Y^\dag=YX^\dag C_U(\rho)$ contains a unitary tensor decomposable
element $U_1\otimes U_2$.

Conversely, assume $GXY^\dag$ contains a tensor decomposable
element $U_1\otimes U_2$. We have then $UXY^\dag=U_1\otimes U_2$,
$U\rho=\rho U$ and $U$ is unitary. Therefore
$$\ba{rcl}
\rho^\prime&=&Y\Lambda Y^\dag=(U_1\otimes U_2)^\dag U X
\Lambda X^\dag U^\dag (U_1\otimes U_2)\\
&=&(U_1\otimes U_2)^\dag U \rho U^\dag (U_1\otimes U_2)
=(U_1\otimes U_2)^\dag \rho (U_1\otimes U_2).
\ea
$$
\hfill $\Box$

Let $Z$ be an $M\times M$ block matrix with each block of size
$N\times N$, the realigned matrix $\tilde{Z}$ is defined by
$$
\tilde{Z}=[vec(Z_{11}),\cdots,vec(Z_{M1}),\cdots,
vec(Z_{1M}),\cdots, vec(Z_{MM})]^t,
$$
where for any $M\times N$ matrix $A$  with  entries $a_{ij}$,
$vec(A)$ is defined to be
$$
vec(A)=[a_{11},\cdots,a_{M1},a_{12},\cdots,a_{M2},\cdots,a_{1N},\cdots,
a_{MN}]^t.
$$
It is verified that a matrix $V$ can
be expressed as the tensor product of two matrices
$V_1$ and $V_2$, $V=V_1\otimes V_2$  if  and  only  if (cf, e.g., \cite{kropro})
$\tilde{V}=vec(V_1)vec(V_2)^t$.
Moreover \cite{tri}, for an $MN\times MN$ unitary matrix $U$,
if $U$ is a unitary decomposable matrix, then the rank of
$\tilde{U}$ is one,  $r(\tilde{U})=1$.
Conversely if $r(\tilde{U})=1$, there exists an $M\times M$ matrix
$U_1$ and an $N\times N$ matrix $U_2$, such that $U=U_1\otimes U_2$ and
$U_1U_1^{\dag}=U_1^{\dag}U_1=k^{-1}I_M$, $U_2U_2^{\dag}=U_2^{\dag}U_2=kI_N$,
where $I_N$ (resp. $I_M$) denotes the $N\times N$ (resp. $M\times M$)
identity matrix, $k>0$, and $U$ is a unitary tensor decomposable matrix.

{\sf [Theorem].}
Let $\rho$ and $\rho^\prime$ be two density matrices
with orthonormal unitary matrices $X$ and $Y$ as given in (\ref{xy}).
If $\rho$ and $\rho^\prime$ are not degenerate,
they are equivalent under local unitary transformations if and only if the set of
matrices $XDY^\dag$, $D=diag(e^{i\theta_1},\,e^{i\theta_2},...,e^{i\theta_{MN}})$,
contains a unitary tensor decomposable element for some $\theta_i\in \Rb$.

{\sf [Proof].}
For each $g\in G=C_U(\rho)$, we have $g\rho=\rho g$, or
$$
gX\left(
\ba{cccc}
\lambda_1I_{n_1}&0&\cdots&0\\
0&\lambda_2I_{n_2}&\cdots&0\\
\vdots&&\ddots&\vdots\\
0&\cdots&\cdots&\lambda_rI_{n_r}
\ea
\right)X^\dag
=X\left(
\ba{cccc}
\lambda_1I_{n_1}&0&\cdots&0\\
0&\lambda_2I_{n_2}&\cdots&0\\
\vdots&&\ddots&\vdots\\
0&\cdots&\cdots&\lambda_rI_{n_r}
\ea
\right)X^\dag g.
$$
That is, $X^\dag g X$ commutes with the diagonal block matrix
$diag(\lambda_1I_{n_1},\lambda_2I_{n_2},...,\lambda_rI_{n_r})$.
Therefore $X^\dag g X$ has the form
\be\label{xgx}
X^\dag g X=diag(A_{n_1},A_{n_2},...,A_{n_r}),
\ee
where $n_i$, $i=1,2,...,r$, stands for the geometric multiplicity of the eigenvalue $\lambda_i$
of $\rho$, $\sum_1^r n_r=MN$, $A_{n_i}$ are $n_i\times n_i$ complex matrices.
As $X^\dag g X$ is unitary, $A_{n_i}$ are also unitary.

For the case that $\rho$ and $\rho^\prime$ are not degenerate, the
$n_i\times n_i$ unitary matrices $A_{n_i}$ become phase factors
$A_{n_i}=e^{i\theta_i}$, as in this case $n_i=1$.
Set $V=gXY^\dag$. From (\ref{xgx}) we have
\be\label{v}
V=XDY^\dag,~~~~D=diag(e^{i\theta_1},\,e^{i\theta_2},...,e^{i\theta_{MN}}).
\ee
According to the {\sf Lemma} $\rho$ and $\rho^\prime$ are equivalent under
local unitary transformations if and only if $V$ is unitary tensor
decomposable, i.e. rank of the realigned matrix $\tilde{V}$ is one, $r(\tilde{V})=1$.
\hfill $\Box$

The {\sf Theorem} presents an operational way to verify whether two non-degenerate
bipartite mixed states $\rho$ and $\rho^\prime$ are equivalent or
not under local unitary transformations. One only needs to
calculate the matrices $X$ and $Y$ in (\ref{xy}) by calculating
their orthonormal eigenvectors, and check if the rank of the
realigned matrix $\tilde{V}$ of (\ref{v}) could be one. If $r(\tilde{V})=1$, one gets $U_1$
and $U_2$ such that $V=U_1\otimes U_2$, and $\rho$,
$\rho^\prime$ are equivalent under local unitary transformations.

As an example we consider a density matrix on $2\times 4$,
\be\label{exam}
\rho =
\left(\ba{cccccccc}1& 0& 0& 0& 0& 0& 0& 1\\0& a& 0& 0& 0& 0& 0& 0\\0& 0& b& 0& 0& 0&
       0& 0\\0& 0& 0& c& 0& 0& 0& 0\\0& 0& 0& 0& 1/c& 0& 0& 0\\0& 0& 0&
      0& 0& 1/b& 0& 0\\0& 0& 0& 0& 0& 0& 1/a& 0\\1& 0& 0& 0& 0& 0& 0& 1\ea\right),
\ee
which is in fact a PPT entangled edge state if it is viewed as a three-qubit state
\cite{prl}.
Let's consider $\rho^\prime$ to be another state of the form (\ref{exam})
with $a\,,b\,,c$ replaced by $a^\prime=b$, $b^\prime=c$, $c^\prime=a$.
Then $\rho^\prime$ and $\rho$ have the same eigenvalue set.
Calculating the unitary matrices $X$ and $Y$ that diagonalize $\rho$ and
$\rho^\prime$, we get $\rho=X\Lambda X^\dag$, $\rho^\prime=Y\Lambda Y^\dag$, where
$\Lambda=diag(2,0,1/a,a,1/b,b,1/c,c)$. For the case $a\neq b\neq c\neq 1$, $2$ and $1/2$,
$\rho$ and $\rho^\prime$ are not degenerate. In this case we have
the matrix $V$ defined by (\ref{v}),
$$
V=\left(
\ba{cccccccc}
d_1 + d_8& -d_1 + d_8& 0& 0& 0& 0& 0& 0\\-d_1 + d_8& d_1 + d_8& 0& 0& 0& 0& 0&
    0\\0& 0& 0& 0& d_7& 0& 0& 0\\0& 0& 0& 0& 0& d_2& 0& 0\\0& 0& 0& 0& 0&
    0& d_6& 0\\0& 0& 0& 0& 0& 0& 0& d_3\\0& 0& d_5& 0& 0& 0& 0& 0\\0& 0& 0&
     d_4& 0& 0& 0& 0
\ea\right).
$$
Therefore
$$
\tilde{V}=\left(
\ba{cccccccccccccccc}
d_1 + d_8& -d_1 + d_8& 0& 0& -d_1 + d_8& d_1 + d_8& 0& 0& 0& 0& 0& 0& 0& 0& 0& 0\\
0& 0& d_7& 0& 0& 0& 0& d_2& 0& 0& 0& 0& 0& 0& 0& 0\\
0& 0& 0& 0& 0& 0& 0& 0& 0& 0& d_5& 0& 0& 0& 0& d_4\\
0& 0& 0& 0& 0& 0& 0& 0& d_6& 0& 0& 0& 0& d_3& 0& 0
\ea\right),
$$
which can not be a rank one matrix for $d_i=e^{i\theta_i}$. Hence $\rho$ and
$\rho^\prime$ are not equivalent under local unitary transformations.

The approach presented in this paper is rather different from the ones in
\cite{Rains,Grassl,makhlin,linden,generic,goswami,tri}, where a set of invariants
has to be calculated for some classes of quantum states. In fact
our approach works for all bipartite mixed states. But
such criterion is dramatically simplified
when the degeneracy of the related density matrices is
reduced. In particular, for the non-degenerate case, two density matrices are
easily verified whether they are equivalent or not under
local unitary transformation (as $e^{i\theta_j}\neq 0$, one can effectively
compute the rank of the realigned matrix of $V$ in (\ref{v}) and see
whether it is possible for the rank to be one).

The results above can be also used to verify the equivalence of pure multipartite
states \cite{new1}. For instance, we can consider a pure tripartite state $\rho_{ABC}$ with
subsystems, say, $A,B$ and $C$ as bipartite states $\rho_{A|BC}$,
$\rho_{AB|C}$ or $\rho_{AC|B}$. If for one of the bipartite decompositions, say
$\rho_{A|BC}$, we have $\rho_{A|BC}^\prime=(U_A\otimes U_{BC})\rho_{A|BC}(U_A\otimes U_{BC})^{\dag}$,
we consider further
$\rho_{BC}=Tr_A (\rho_{A|BC})$, which is a mixed bipartite state
and can be judged by using our theorems. If
$\rho_{BC}^\prime=(U_B\otimes U_{C})\rho_{BC}(U_B\otimes U_{C})^{\dag}$,
we have
$\rho_{ABC}^\prime=(U_A\otimes U_{B}\otimes U_{C})\rho_{ABC}(U_A\otimes U_{B}\otimes U_{C})^{\dag}$.
In this way the equivalence for a class of pure
tripartite states can also be studied according to our theorems.

\vspace{0.8truecm}
\noindent {\bf Acknowledgments} S.M. Fei
gratefully acknowledges the warm hospitality of Dept. Math. in NCSU and
the support provided by American Mathematical Society.
The work is partially supported by a NKBRPC (2004CB318000).

\end{document}